\documentclass[aps,twocolumn,pra,reprint,amsmath,amssymb,floatfix,footinbib,superscriptaddress]{revtex4}

\usepackage{amsmath}	
\usepackage{gensymb}
\usepackage{hyperref}
\hypersetup{colorlinks=true}
\usepackage{upgreek}
\usepackage{graphics}
\usepackage{hyperref}
\usepackage{epsfig}
\usepackage{color}
\usepackage{bm}
\usepackage{float}
\usepackage{ulem} 
\usepackage{blindtext}

\usepackage{CJK}        
\usepackage{url}
\usepackage{multirow}

\newcommand{\upperRomannumeral}[1]{\uppercase\expandafter{\romannumeral#1}}

\begin{document}

\title{Analytical results for the unusual Gr{\"u}neisen ratio \\ in the quantum Ising model with Dzyaloshinskii-Moriya interaction}

\author{Qiang Luo\,\href{https://orcid.org/0000-0001-8521-0821}{\includegraphics[scale=0.12]{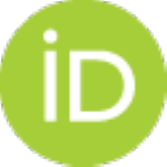}}}
\email[]{qiangluo@nuaa.edu.cn}
\affiliation{College of Science, Nanjing University of Aeronautics and Astronautics, Nanjing, 211106, China}
\affiliation{Key Laboratory of Aerospace Information Materials and Physics (NUAA), MIIT, Nanjing, 211106, China}

\date{\today}

\begin{abstract}
  The Gr{\"u}neisen ratio (GR) has emerged as a superb tool for the diagnosis of quantum phase transitions,
  which diverges algebraically upon approaching critical points of continuous phase transitions.
  However, this paradigm has been challenged recently by observations of a finite GR for self-dual criticality
  and divergent GR at symmetry-enhanced first-order transitions.
  To unveil the fascinating GR further, we exemplify the idea by studying an exactly solvable quantum Ising model with Dzyaloshinskii-Moriya interaction,
  which harbors a ferromagnetic phase, a paramagnetic phase, and a chiral Luttinger liquid.
  Although the self-dual criticality of the ferromagnetic--paramagnetic transition is undermined by the Dzyaloshinskii-Moriya interaction,
  we find that the GR at the transition is still finite albeit with an increasing value,
  signifying a proximate self-dual relation.
  By contrast, the GR at the transition between the gapped ferromagnetic phase and the gapless Luttinger liquid diverges and changes its sign
  when crossing the first-order transition.
  This implies that the GR could also probe the first-order transition between the gapped and gapless phases.
\end{abstract}


\maketitle

\textit{Introduction}.---
The quantum phase transitions~(QPTs) are ubiquitous phenomena which occur upon tuning external parameters in the lowest temperature \cite{Vojta2003,RossiniVicari2021}.
They are accompanied by singular changes of the ground state which could be probed by information-theoretic quantities,
for instance, the entanglements \cite{Osterloh2002,OsborneNielsen2002,VLRK2003,AmicoRMP2008} and fidelity susceptibility \cite{ZanardiPaun2006,YouLiGu2007,CamposVenuti2007}.
As temperature increases, the interplay between thermal and quantum fluctuations strongly promotes the formation of a quantum critical region of continuous QPTs.
Physically, this special area is always of keen interest as it is beneficial for unconventional events,
such as non-Fermi liquid behavior in metals \cite{Chowdhury2018}, deconfined quantum criticality \cite{Senthil2004}, and superconductivity \cite{MarelHTSC2003,ZhouHTSC2021}.
Therefore, the thermodynamic quantities, which can not only diagnose the QPTs but are also experimentally accessible, are highly desirable.

So far, the Gr{\"u}neisen ratio~(GR) $\Gamma$ \cite{ZGRS2003,GarstRosch2005},
which is defined as the ratio between magnetic expansion coefficient $\alpha_T$ and specific heat $C_v$,
is extremely remarkable.
It is shown that the GR diverges algebraically at the QCP as $\Gamma \sim T^{-1/(\nu z)}$
(Here, $z$ and $\nu$ are the dynamical and correlation length critical exponents, respectively.) \cite{ZGRS2003},
and also undergoes a sign change in the vicinity of the QCP \cite{GarstRosch2005}.
Such an abnormal behavior of GR has been actively studied in many quantum systems including
strongly interacting quantum gases \cite{DeSouza2016,PengYuGuan2019,YuZhangGuan2020}, itinerant electron systems \cite{WatanabeMiyake2019},
and quantum spin models \cite{Jafari2012,You2014a,GomesSqu2019}.
Over the years, the GR has been used to identify and characterize QPTs in various materials (for review see Ref.~\cite{Gegenwart2017}),
including heavy-fermion systems \cite{Kuchler2003,Kuchler2004,Kuchler2006,Tokiwa2009,Tokiwa2015},
spin-chain material BaCo$_2$V$_2$O$_8$ \cite{WangBCVO2018},
and spin-liquid candidate $\alpha$-RuCl$_3$ \cite{BachusKT2020}.

However, the divergence of the GR at QCPs is not a universal character of all continuous QPTs.
Based on the hyperscaling theory, it is demonstrated by Zhang that the GR remains finite for self-dual QCPs \cite{Zhang2019}.
The quantum Ising model $\mathcal{H}(g)$ with $g$ being the external magnetic field is perhaps the most prominent example which exhibits the self-duality.
This model owns an intrinsic $\mathbb{Z}_2$ symmetry and can be recast as $g\mathcal{H}(1/g)$.
As such $g_c = 1$ is identified as a self-dual QCP where the GR equals to $1/2$ \cite{Zhang2019,WuZhuSi2018,ZhangDing2019}.
On the other hand, the GR may also diverge at first-order QPTs on certain conditions.
A recent study by Beneke and Vojta shows that in the symmetry-enhanced first-order QPT which is accompanied by a vanishing mode gap,
the GR indeed diverges albeit with mean-field critical exponents \cite{BenekeVojta2021}.
These exceptions imply that a comprehensive understanding of the GR near the QCPs is still lacking,
and it is intriguing to know its behaviors in the proximity of self-dual quantum criticality
and in other first-order QPTs where the energy gaps vanish.

In this work, we demonstrate unusual behaviors of the GR by the quantum Ising model with Dzyaloshinskii-Moriya~(DM) interaction \cite{Jedrzejewski2008}.
This model has an intimate relation to Ising-like spin-1/2 chain compounds BaCo$_2$V$_2$O$_8$ \cite{Faure2018,Wang2018,Wang2019},
SrCo$_2$V$_2$O$_8$ \cite{Okutani2015,Cui2019}, and CoNb$_2$O$_6$ \cite{ColdeaSCI2010,Liang2015,Amelin2020,Morris2021}.
The transverse field is then involved by applying a magnetic field normal to the Ising spin direction.
Meanwhile, for the spin-orbit-coupled bosons embedded in 1D optical lattices,
it can be effectively regarded as an Ising ferromagnet subjected to DM interaction along the $z$ direction
when the intraspecies interaction strength is prominently larger than that of the interspecies one \cite{ZhaoPRA2014,Xu2014,Piraud2014,ZhaoPRB2014,Peotta2014,Xi2017}.
This model can be solved exactly by Jordan-Wigner transformation and it is known to host a ferromagnetic (FM) phase, a paramagnetic (PM) phase,
and a Luttinger liquid with chiral ordering (hereafter termed chiral-LL) \cite{Derzhko2006,Soltania2019,Zhong2019,DingZhong2021,You2014b}.
As will be demonstrated analytically below, the GR in the model exhibits abnormal behaviors.
On the one hand, although the FM--PM transition is still continuous, the exact self-dual relation is ruined when the DM interaction is involved.
However, the GR at the critical point remains finite, albeit with a tendency to diverge as the DM interaction increases.
The nondivergence of the GR herein comes from the self-duality rooted in the quantum Ising model and the peculiarity of the DM interaction.
On the other hand, while the transition between the gapped FM phase and the gapless chiral-LL phase is of first order,
the energy gap vanishes at the transition and the GR displays a typical power-law divergence as that of a continuous QPT.

\begin{figure}[!ht]
\centering
\includegraphics[width=0.95\columnwidth, clip]{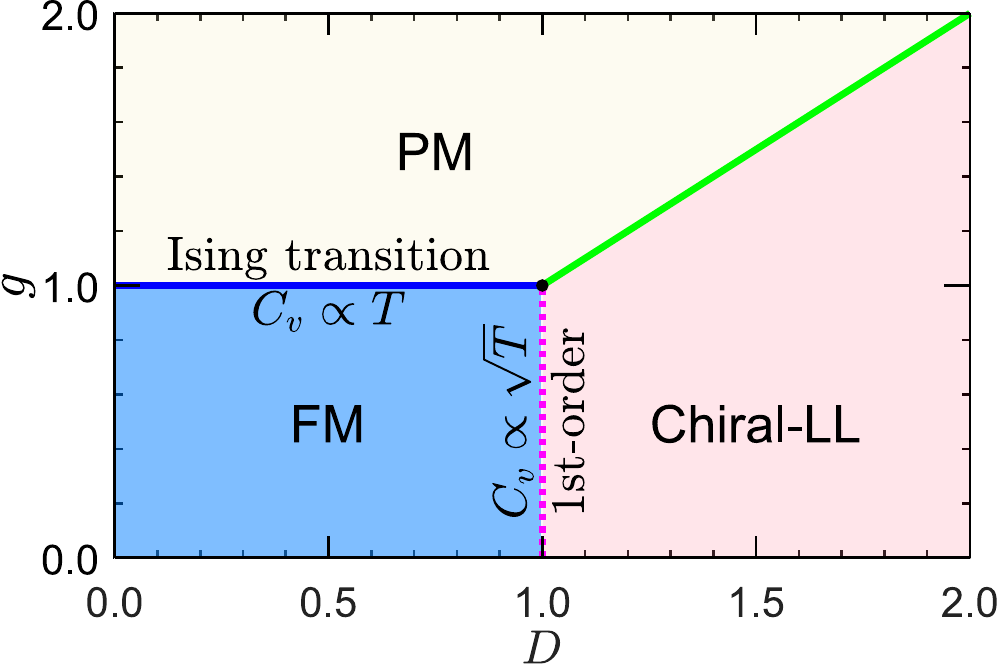}\\
\caption{The phase diagram of the quantum Ising model with DM interaction.
  There are three distinct phases, which are a gapped FM phase (blue), a gapped PM phase (gray), and a gapless chiral-LL phase (pink).
  The FM--PM transition is continuous of the Ising type, while the FM--chiral-LL transition is of first order.
  The low-temperature specific heats $C_v$ on the critical lines exhibit the behaviors of $C_v \propto T$ and $C_v \propto \sqrt{T}$, respectively.
  }\label{FIG-QIMPD}
\end{figure}

\textit{Model}.---
The Hamiltonian of the quantum Ising model with DM interaction can be described by \cite{Jedrzejewski2008}
\begin{align}\label{EQ:HamQIMvDMI}
\mathcal{H} \!=\! -\sum_{j=1}^{N} \Big[\sigma_j^x\sigma_{j+1}^x + g\sigma_j^z + \frac{D}{2}(\sigma_j^x\sigma_{j+1}^y\!-\!\sigma_j^y\sigma_{j+1}^x)\Big],
\end{align}
where $\sigma_j^{\gamma}$~($\gamma$ = $x$, $y$, and $z$) is the $\gamma$-component of the Pauli operator acting on site $i$,
$g$ is the magnetic field, and $D$ denotes the strength of the DM interaction along the $z$ direction.
Following the standard prescription, this model can be diagonalized analytically via the Jordan-Wigner transformation \cite{Pfeuty1970},
\begin{align}\label{EQ:JWTran}
\sigma_j^x &= \prod_{m<j} (1-2c_m^{\dagger}c_m) (c_j + c_j^{\dagger}), \nonumber\\
\sigma_j^y &= -\imath \prod_{m<j} (1-2c_m^{\dagger}c_m) (c_j - c_j^{\dagger}), \\
\sigma_j^z &= 1-2c_j^{\dagger}c_j, \nonumber
\end{align}
which maps spins into spinless fermions with creation (annihilation) operators $c_j^{\dagger}$ ($c_j$).
By transforming the spinless fermion operator to momentum space $c_j = \frac{1}{\sqrt{N}} \sum_k c_k e^{\imath 2\pi jk/N}$
and exploiting the Bogoliubov transformation $c_k = u_k\gamma_k + \imath v_k\gamma_{-k}^{\dagger}$
where $u_k = \cos(\theta_k/2)$ and $v_k = \sin(\theta_k/2)$ with $\tan\theta_k = \sin k/(g - \cos k)$,
the Hamiltonian \eqref{EQ:HamQIMvDMI} can be recast into the following form
\begin{equation}\label{EQ:HamBogoliu}
\mathcal{H} = \sum_{k} \epsilon_k \left(\gamma_k^{\dagger}\gamma_k - \frac12\right),
\end{equation}
where the dispersion relation reads
\begin{equation}\label{EQ:EgDisp}
\epsilon_k = 2\big(\sqrt{1+g^2-2g\cos k} + D\sin k\big).
\end{equation}
For the discussion on the occurrence of the Fermi points in the reciprocal space, see Sec. S1 in the Supplemental Material (SM) \cite{SuppMat}.
The free energy density is calculated as
\begin{equation}\label{EQ:FreeEgDef}
F(T) = -\frac{1}{\beta} \left[\ln 2 + \frac{1}{2\pi}\int_{-\pi}^{\pi} dk \ln \cosh \Big(\frac{\beta\epsilon_k}{2}\Big)\right]
\end{equation}
where $\beta = 1/k_BT$ (hereafter $k_B$ = 1).
The thermodynamic quantities such as the thermal entropy $S$ and specific heat $C_v$ can be obtained via Eq.~\eqref{EQ:FreeEgDef}.
For example, the specific heat is given by
\begin{equation}\label{EQ:CvHc}
C_v = \frac{1}{2\pi}\int_{-\pi}^{\pi} dk \Big(\frac{\beta\epsilon_k}{2}\Big)^2 \textrm{sech}^2\Big(\frac{\beta\epsilon_k}{2}\Big).
\end{equation}
In this context, the GR $\Gamma(T, \lambda)$ is defined as \cite{ZGRS2003,GarstRosch2005}
\begin{equation}\label{EQ:GRDef}
\Gamma(T, \lambda) = \frac{1}{T}\left(\frac{dT}{d\lambda}\right)_S
= - \frac{(\partial S/\partial\lambda)_T}{T(\partial S/\partial T)_{\lambda}}
= - \frac{\alpha_T}{C_v}
\end{equation}
where $\alpha_T = (\partial S/\partial\lambda)_T$ is the magnetic expansion coefficient,
with $\lambda$ ($g$ and $D$) being a tuneable parameter.

\begin{figure}[!ht]
\centering
\includegraphics[width=0.90\columnwidth, clip]{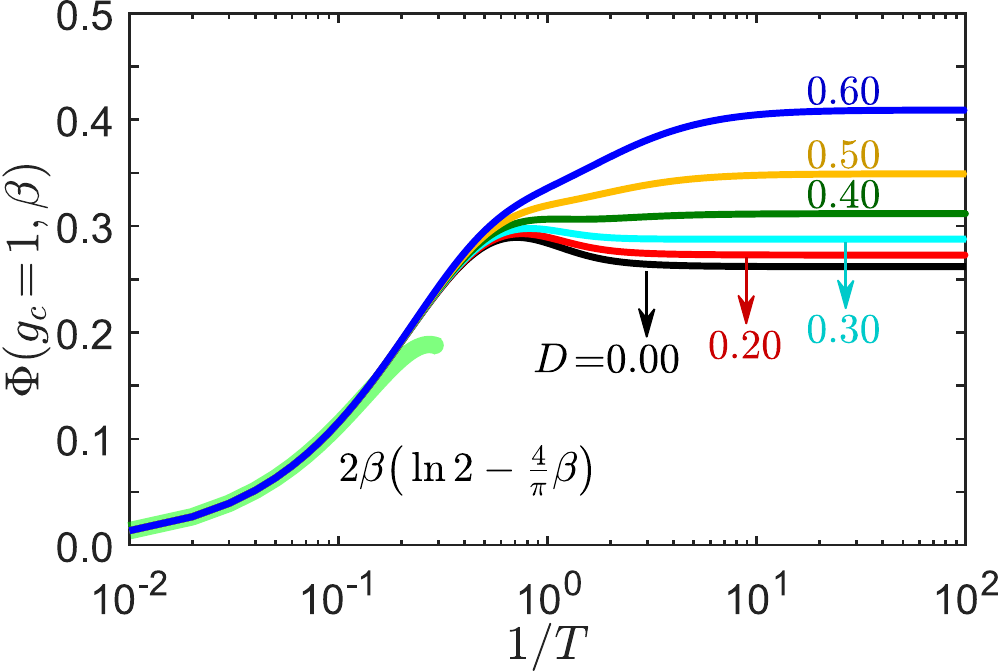}\\
\caption{The scaled free energy $\Phi_g(T, D)$ along the Ising transition line with $g_c = 1$.
  In the high-$T$ region, it is insensitive to $D$ (as marked by a green belt),
  while in the low-$T$ region it saturates to constant values of 0.2618 ($D$ = 0.0, black), 0.2727 ($D$ = 0.2, red), 0.2877 ($D$ = 0.3, cyan),
  0.3117 ($D$ = 0.4, green), 0.3491 ($D$ = 0.5, yellow), and 0.4091 ($D$ = 0.6, blue), respectively.
  }\label{FIG-FreeEgHz}
\end{figure}

\begin{figure*}[htb]
\centering
\includegraphics[width=1.65\columnwidth, clip]{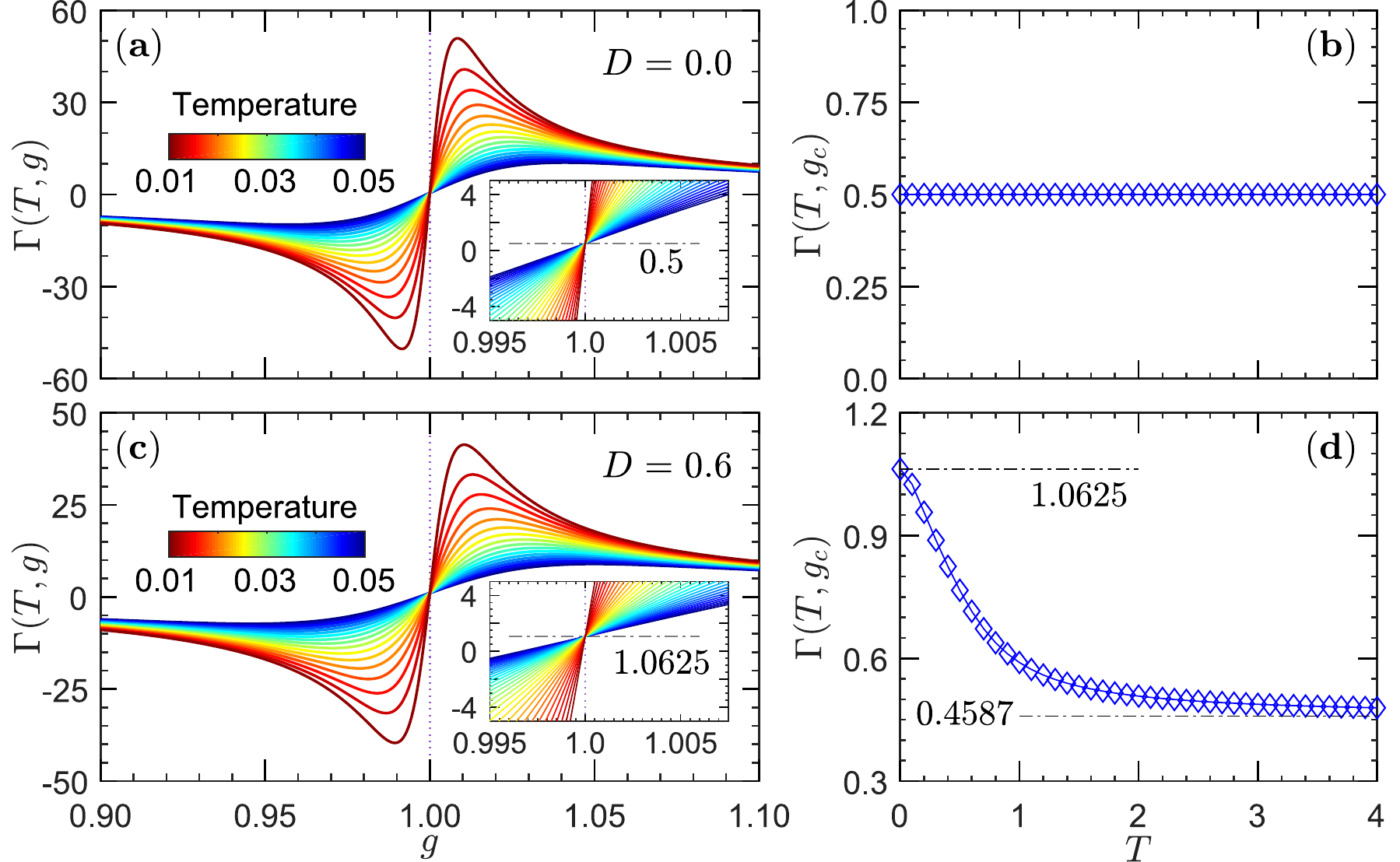}\\
\caption{(a) The magnetic field $g$ dependence of the GR $\Gamma(T, g)$ in the temperature interval of [0.01, 0.05] with DM interaction $D$ = 0.0.
  The inset shows the GR in the vicinity of $g_c = 1$, where different curves intersect with a value of 0.5 \textit{exactly}.
  (b) The GR $\Gamma(T, g_c = 1)$ plotted as a function of $T$ with $D$ = 0.0. It is a constant of 0.5 that is irrelevant of temperature.
  The panels (c) and (d) are respectively analogs of (a) and (b) but for $D$ = 0.6.
  Inset of (c) shows the GR in the vicinity of $g_c = 1$, where different curves intersect with a value of 1.0625 \textit{asymptotically}.
  (d) As $T$ decreases to the lowest temperature, the GR $\Gamma(T, g_c = 1)$ increases from 0.4587 (when $T\to\infty$) to 1.0625 (when $T\to0$).
  Behaviors of the GR $\Gamma(T\to0, g)$ in the zero-temperature limit are shown in Sec. S4 in the SM \cite{SuppMat}.
  }\label{FIG-GRDz000}
\end{figure*}

To illuminate the overview of the model, we recap the main features of the phase transitions in Fig.~\ref{FIG-QIMPD}.
There are three distinct phases which are known as the FM phase when $g$ and $D$ are small,
the PM phase when the magnetic field $g$ is very strong, and the chiral-LL phase in the presence of large DM interaction.
The FM--PM transition is recognized to belong to the Ising universality class
with both critical exponents $\nu$ and $z$ being 1.
By contrast, the transition between the FM phase and the chiral-LL phase is of first order but with a vanishing energy gap in the transition line.
We also note that the FM--chiral-LL transition, which occurs at the line of $g = D$ ($g, D > 1$), is continuous and is a possible realization of Dzhaparidze-Nersesyan-Pokrovsky-Talapov universality class \cite{DzhNer1978,PokTal1979}.
However, identifying this transition type is beyond the scope of the current work.

\textit{Proximate self-dual criticality}.---
To study the quantum criticality in a relevant temperature region,
it is useful to recall the so-called scaled free energy coefficient introduced by Kopp and Chakravarty \cite{KoppChak2005}.
For the Ising transition line where $g_c = 1$, it is found that
\begin{equation}\label{EQ:PhiGDef}
\Phi_g(T, D) = \frac{2}{T^2} \big(F(0) - F(T)\big)
\end{equation}
where the ground-state energy $F(0)$ equals to $-4/\pi$ regardless of the value of $D$.
Figure~\ref{FIG-FreeEgHz} shows the behavior of $\Phi_g(T)$ at several different values of $D$.
In the high-temperature region, $\Phi_g(T, D)$ is insensitive to $D$ and obeys the rule of $2\beta\big(\ln2 - \frac{4}{\pi}\beta\big)$ approximately.
With the temperature decreasing, $\Phi_g(T, D)$ displays a broad hump and then saturates to a constant below a character temperature $T^{*}$.
Physically, this constant is identical to the specific-heat coefficient at the low temperature,
and the value of $T^{*}$ denotes the upper range of the quantum critical scaling region.
When $D = 0$, it is well known that $\Phi_g(T\to0) = \pi/12$ \cite{BloteCN1986,Affleck1986} and $T^{*} \simeq 1/2$,
showing that the quantum criticality can persist up to a temperature which is as large as one half of the energy unit.
However, in the presence of DM interaction, we find that the plateau of $\Phi_g(T, D)$ grows rapidly while the $T^{*}$ is suppressed with the increase of $D$.
As demonstrated in Sec. S2 in the SM \cite{SuppMat}, we find that $\Phi_g(T\to0, D)$ = $\pi/[12(1-D^2)]$.

As the magnetic field is away from the critical point, there is a continuous QPT which belongs to the Ising universality class.
For this transition, the GR can be calculated via Eq.~\eqref{EQ:GRDef} where the magnetic expansion coefficient
\begin{equation}\label{EQ:MagExpHc}
\alpha_g = \frac{1}{2\pi}\int_{-\pi}^{\pi} dk \frac{\beta^2\epsilon_k(g-\cos k)}{2\sqrt{1+g^2-2g\cos k}} \textrm{sech}^2\Big(\frac{\beta\epsilon_k}{2}\Big).
\end{equation}
The magnetic-field dependence of the GR $\Gamma(T, g)$ in the temperature interval of [0.01, 0.05] is plotted
in Fig.~\ref{FIG-GRDz000}(a) and (c) with DM interaction $D$ = 0.0 and 0.6, respectively.
In panel (a), the critical point is demonstrated to own a self-dual relation, resulting in a nondivergent GR.
Crucially, as shown in the inset and also in Fig.~\ref{FIG-GRDz000}(b),
the GR intersects precisely at the QCP with a value of $1/2$ that is irrelevant of temperature.
On the other hand, while the GR shown in panel (c) is also finite in the QCP,
it is nonconstant and varies with the temperature.
As $T$ decreases to the lowest temperature, the GR $\Gamma(T, g_c = 1)$ increases
from 0.4587 (when $T\to\infty$) to 1.0625 (when $T\to0$), see Fig.~\ref{FIG-GRDz000}(d).

The unusual temperature dependence of the GR at the QCP can be understood analytically. When $g$ = 1, we find that the GR can be simplified as
\begin{equation}\label{EQ:GruneisenRatio}
\Gamma(T, g_c) = -\frac{\alpha_g}{C_v} = \frac{\mathcal{I}_1(T, D) + \mathcal{I}_1(T, -D)}{\mathcal{I}_2(T, D) + \mathcal{I}_2(T, -D)}
\end{equation}
where
\begin{align}\label{EQ:FuncInu}
\mathcal{I}_{\upsilon}(T, D)
=& \frac{\upsilon}{\pi} \int_0^{\pi/2} dk (1+D\cos k)^{\upsilon} (2\beta\sin k)^2 \nonumber\\
 & \times \textrm{sech}^2 \big[2\beta\sin k (1+D\cos k)\big].
\end{align}
In the high-temperature limit where $T\to\infty$ (i.e., $\beta\to0$), we have
$\mathcal{I}_1 \simeq \beta^2\left[1 + {4D}/{(3\pi)}\right]$
and
$\mathcal{I}_2 \simeq \beta^2\left[{16D}/{(3\pi)} + {(4+D^2)}/{2}\right]$.
Therefore, the GR behaves as
\begin{equation}\label{EQ:GRTtoinfty}
\Gamma(T\to\infty, g_c) = \frac{2}{4+D^2} \leq \frac{1}{2}.
\end{equation}
In the low-temperature limit where $T\to0$ (i.e., $\beta\to\infty$), we have
$\mathcal{I}_1 \simeq 1/{[2\pi\beta(1+D)^2]}$
and
$\mathcal{I}_2 \simeq 1/{[\pi\beta(1+D)]}$.
The GR is
\begin{equation}\label{EQ:GRTtoZero}
\Gamma(T\to0, g_c) = \frac{(1+D^2)}{2(1-D^2)} \geq \frac{1}{2}.
\end{equation}
The lower limit and the upper limit of the GR revealed in Eq.~\eqref{EQ:GRTtoinfty} and Eq.~\eqref{EQ:GRTtoZero} are plotted in Fig.~\ref{FIG-GRDz000}(d) with $D$ = 0.6.
The curve in Fig.~\ref{FIG-GRDz000}(d) may be beneficial for extracting exchange parameters in real materials like BaCo$_2$V$_2$O$_8$ \cite{WangBCVO2018}.
It is inferred from Eq.~\eqref{EQ:GRTtoZero} that the GR will diverge in the case of $D = 1$ where the conventional continuous QPT occurs.
Therefore, our study reveals the precious evolution of the GR when away from the self-dual QCP.

\begin{figure}[!ht]
\centering
\includegraphics[width=0.90\columnwidth, clip]{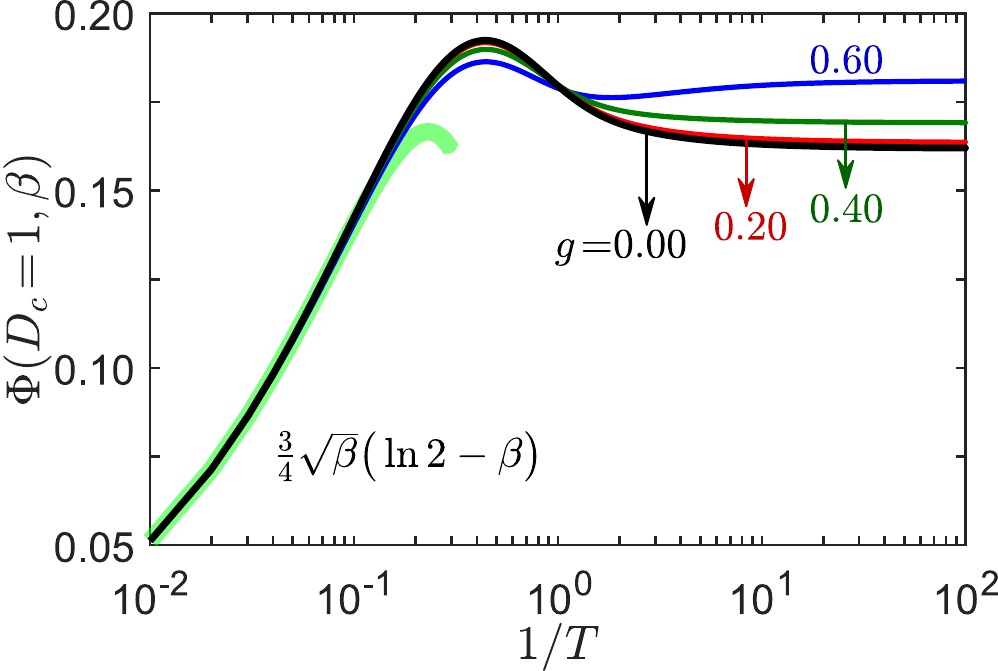}\\
\caption{The scaled free energy $\Phi_D(T, g)$ along the first-order transition line with $D_c = 1$.
  In the high-$T$ region, it is insensitive to $g$ (as marked by a green belt),
  while in the low-$T$ region it saturates to constant values of 0.1619 ($g$ = 0.0, black), 0.1636 ($g$ = 0.2, red),
  0.1691 ($g$ = 0.4, green), and 0.1809 ($g$ = 0.6, blue), respectively.
  }\label{FIG-FreeEgDM}
\end{figure}

\begin{figure*}[htb]
\centering
\includegraphics[width=1.65\columnwidth, clip]{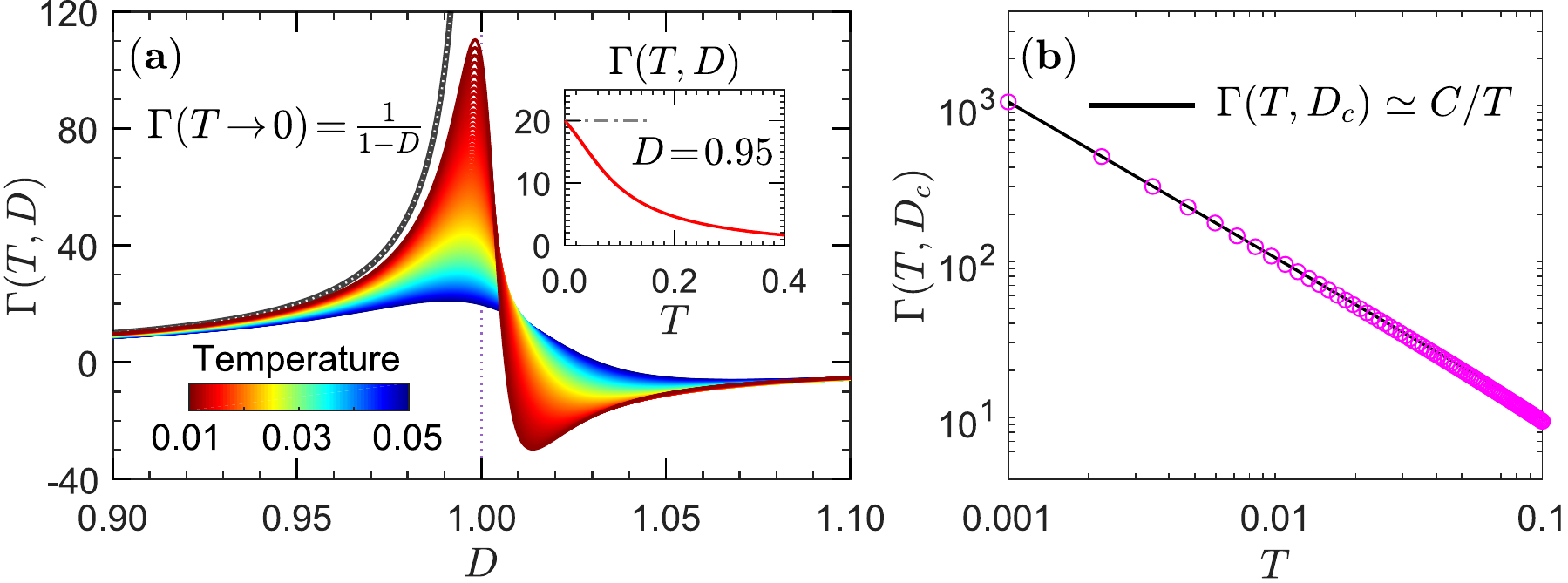}\\
\caption{(a) The DM interaction $D$ dependence of the GR $\Gamma(T, D)$ in the temperature interval of [0.01, 0.05] with magnetic field $g$ = 0.0.
  The black line of $1/(1-D)$ represents the asymptotic behavior of $\Gamma(T\to0, D)$.
  Inset: Evolution of $\Gamma(T, D)$ as a function of $T$ at fixed $D$ = 0.95.
  (b) The GR $\Gamma(T, D_c = 1)$ plotted as a function of $T$ with $g$ = 0.0.
  The pink open circles are the selected data while the solid black line is the fitting formula with the best fitting constant $C \approx 1.0541$.
  }\label{FIG-GRHz000}
\end{figure*}

In closing, we attempt to uncover the origin of the nondivergence of the GR at the QCP.
According to the hyperscaling ansatz \cite{ZGRS2003}, the GR in the quantum critical region has the form of
$\Gamma(T, g\to g_c) = - G_T T^{-1/(z\nu)}$ where $G_T \propto \Psi'(0)$ (for details, see Sec. S3 in the SM \cite{SuppMat}).
Here, $\Psi'(0)$ is the linear scaling term in the reduced free energy $F(T)/T^2$ with respect to $(g-g_c)/T$.
Following the recipe depicted in Ref.~\onlinecite{WuZhuSi2018},
we find that the free energy near the QCP in the low-temperature limit reads
\begin{align}\label{EQ:FreeEgLowTFnl}
F(T) = -\frac{T^2}{\pi} \Big[a_{g,D} + \frac12 \Big(\frac{1-g}{T}\Big)^2\Big]
\end{align}
where $a_{g,D} = \ln2 + (g+D^2/4)/6$.
It is inferred from Eq.~\eqref{EQ:FreeEgLowTFnl} that the linear scaling term of $(1-g)/T$ indeed vanishes and thus $\Psi'(0) = 0$ for arbitrary $D$.
Therefore, the inclusion of the DM interaction does not bring about an extra relevant term but only has a correlation on the constant term
\footnote{Our work demonstrates that the nondivergence of the GR does not necessarily signify an exact self-dual relation,
but could also include the case where the self-dual relation is undermined while the condition $\Psi'(0) = 0$ retains.
In this regard, our study extends the conclusion of Ref.~\onlinecite{Zhang2019}.}.

\textit{The gapped-gapless transition}.---
We now turn to study the quantum criticality at the first-order FM--chiral-LL transition.
The FM phase is gapped and its energy gap vanishes as $2\sqrt{1-g^2}(1-D)$ in the vicinity of the transition line $D_c = 1$.
In contrast to Eq.~\eqref{EQ:PhiGDef}, the scaled free energy coefficient in the transition line is found to be
\begin{equation}\label{EQ:PhiDDef}
\Phi_D(T, g) = \frac{3}{4T^{3/2}} \big(F(0) - F(T)\big)
\end{equation}
where $F(0) = -\frac{2(1+g)}{\pi} E\big(\frac{2\sqrt{g}}{1+g}\big)$ is the ground-state energy,
with $E(x)$ being the elliptic integral of the second kind \cite{AbramowitzStegun1972}.
The scaled free energy $\Phi_D(T, g)$ as a function of $T$ for a series of $g$ is illustrated in Fig.~\ref{FIG-FreeEgDM}.
When the temperature is large enough, we have $\Phi_D(T) = \frac{3}{4}\sqrt{\beta}(\ln2 - \beta)$ (see the green belt).
As the temperature decreases, $\Phi_D(T, g)$ approaches to different levels whose values grow with $g$.
When $g$ = 0, the specific heat is calculated to have a square-root low-temperature behavior at the critical point (see Sec. S5.A in the SM \cite{SuppMat}),
\begin{equation}\label{EQ:CvgEQ0}
C_v = \frac{3(\sqrt2-1)\zeta(\frac32)}{8\sqrt{2\pi}} \sqrt{T} \approx 0.1619 \sqrt{T},
\end{equation}
where $\zeta(s)$ stands for the Riemann $\zeta$-function \cite{AbramowitzStegun1972}.
Obviously, the prefactor of the specific heat in Eq.~\eqref{EQ:CvgEQ0} is consistent with the result of $\Phi_D(T, g = 0)$ (black line) shown in Fig.~\ref{FIG-FreeEgDM}
\footnote{We note in passing that if the Pauli operator $\hat{\boldsymbol{\sigma}}$ in the Hamiltonian
is replaced by the spin-$1/2$ operator $\hat{\textbf{S}} = \hat{\boldsymbol{\sigma}}/2$,
the specific heat coefficient should change to $\frac{3}{8}\sqrt{\frac{3\sqrt2-4}{\pi}}\zeta\large(\frac{3}{2}\large)$ $\approx$ 0.2723.
Such a prefactor has been verified by a transfer-matrix renormalization group calculation (see inset of Fig.~1 in Ref.~\onlinecite{Xi2017}).}.
As $g$ increases, the prefactor exhibits a behavior of $0.1619/(1-0.26g^2)$ approximately when $g$ is small.

Without loss of generality, below we study the GR in the case of $g = 0$.
By analogy with the magnetic expansion coefficient shown in Eq.~\eqref{EQ:MagExpHc}, its analogue that is driven by the DM interaction reads
\begin{equation}\label{EQ:MagExpDz}
\alpha_D = \frac{1}{2\pi}\int_{-\pi}^{\pi} dk \frac{\beta^2\epsilon_k \sin k}{2} \textrm{sech}^2\Big(\frac{\beta\epsilon_k}{2}\Big).
\end{equation}
Figure~\ref{FIG-GRHz000}(a) shows the GR $\Gamma(T, D)$ as a function of DM interaction,
and the QPT is manifested by the sign change of $\Gamma(T, D)$ when crossing $D_c = 1$.
In the zero-temperature limit, the GR near the transition point is solely determined by the behavior of energy gap $\Delta(D)$ \cite{BachusKT2020},
\begin{align}
\lim_{T\to0^{+}}\Gamma(T, D) = -\frac{\Delta'(D)}{\Delta(D)},
\end{align}
where the numerator represents the derivative of $\Delta(D)$.
This implies that if the energy gap closes linearly around the transition point (which is indeed the case in the FM phase since $\Delta(D) = 2\sqrt{1-g^2}(1-D)$ when $|1-D| \ll 1$),
the GR should diverge as $\Gamma(T\to0, D) = 1/(1-D)$.
To confirm it, we present the evolution of GR as a function of $T$ at $D = 0.95$ (see inset of Fig.~\ref{FIG-GRHz000}(a)),
whose value becomes 20 as temperature approaches zero.
An alternative way to understand the divergence of GR near the transition point is shown in Sec. S6 in the SM \cite{SuppMat}.

At the critical point, the magnetic expansion coefficient (see Sec. S5.B in the SM \cite{SuppMat})
\begin{equation}\label{EQ:MagExpgEQ0}
\alpha_D = \frac{(\sqrt2-1)\zeta(\frac12)}{2\sqrt{\pi T}}
\end{equation}
in the low-temperature regime. With Eq.~\eqref{EQ:CvgEQ0} and Eq.~\eqref{EQ:MagExpgEQ0} in mind, the GR at the critical point turns out to be
\begin{equation}
\Gamma(T\to0, D_c) = -\frac{4\sqrt2\zeta(\frac12)}{3\zeta(\frac32)} \frac{1}{T} \approx \frac{1.0541}{T},
\end{equation}
which diverges as $\propto 1/T$ when $T\to0$.
Further, the power-law divergence of the GR is also verified numerically in Fig.~\ref{FIG-GRHz000}(b).
Our analysis indicates that the GR can also probe the first-order QPT when the energy gap vanishes algebraically in the vicinity of the transition point.

\textit{Conclusions}--
In this work, we have studied analytically the abnormal behaviors of the GR in the quantum Ising model with DM interaction.
The peculiar feature of the model is that it hosts two distinct QPTs in which one is continuous with a proximate self-dual QCP
while the other is a first-order transition between gapped and gapless phases.
In the continuous FM--PM transition with nonzero DM interaction, the GR increases gradually as the temperature is lowered towards the absolute zero.
However, in the lowest temperature limit the GR remains finite albeit with a tendency to diverge,
which is a reminiscent of the proximate self-dual relation
where the exact self-duality is eroded but the condition $\Psi'(0) = 0$ retains.
By contrast, in the first-order transition driven by DM interaction, the GR can perceive the QPT by changing its sign when crossing the transition point.
Furthermore, akin to a continuous QPT, it also exhibits a power-law singularity at the transition point.
Our analytical results thus constitute a significant contribution of the GR in diagnosing a broad family of QPTs.
They will be useful in detecting characteristic features of magnetocaloric effect in experiments
and can further guide the exploration of field-induced phenomena in real materials.

\begin{acknowledgements}
I would like to thank Chengxiang Ding, Zhidan Wang, Jianda Wu, and Wen-Long You for the discussion on the manuscript.
I am deeply indebted to Shijie Hu, Xiaoqun Wang, Bin Xi, and Jize Zhao for a previous collaboration on a related work \cite{Xi2017}.
I also acknowledge the hotels and nursing staffs in Yancheng for their kind hospitality when I was kept in quarantine.
This work is supported by the startup Fund of Nanjing University of Aeronautics and Astronautics under the Grant No. YAH21129.
\end{acknowledgements}


%




\clearpage

\onecolumngrid

\newpage

\newcounter{sectionSM}
\newcounter{equationSM}
\newcounter{figureSM}
\newcounter{tableSM}
\stepcounter{equationSM}
\setcounter{section}{0}
\setcounter{equation}{0}
\setcounter{figure}{0}
\setcounter{table}{0}
\setcounter{page}{1}
\makeatletter
\renewcommand{\thesection}{\textsc{S}\arabic{section}}
\renewcommand{\theequation}{\textsc{S}\arabic{equation}}
\renewcommand{\thefigure}{\textsc{S}\arabic{figure}}
\renewcommand{\thetable}{\textsc{S}\arabic{table}}


\begin{center}
{\large{\bf Supplemental Material for\\
``Analytical results for the Gr{\"u}neisen ratio in the quantum Ising model with Dzyaloshinskii-Moriya interaction''}}
\end{center}
\begin{center}
Qiang Luo$^{1,\;2}$ \\
\quad\\
$^1$\textit{College of Science, Nanjing University of Aeronautics and Astronautics, Nanjing, 211106, China}\\
$^2$\textit{Key Laboratory of Aerospace Information Materials and Physics (NUAA), MIIT, Nanjing, 211106, China}\\
(Dated: January 20, 2022)
\quad\\
\end{center}


\onecolumngrid

In this supplemental material~(SM), we present the detailed derivation of the Gr{\"u}neisen Ratio~(GR) mentioned in the main text.
It contains six Sections; the first four are about the ferromagnetic-paramagnetic Ising transition,
while the remaining two are about the first-order transition.
Section~\ref{SMSec:I} shows the positions of the Fermi points in the gapless chiral-LL phase,
and In Sec.~\ref{SMSec:II} we present the low- and high-temperature behaviors of GR at the Ising transition.
The nondivergence of the GR in the quantum critical region is analysed in Sec.~\ref{SMSec:III},
while the GR in the zero-temperature limit along the Ising transition line is shown in Sec.~\ref{SMSec:IV}.
In Sec.~\ref{SMSec:V} and Sec.~\ref{SMSec:VI},
we take $g$ = 0 as an example to demonstrate the divergent GR at the first-order transition,
followed by studying the GR in the zero-temperature limit.

\vspace{-0.00cm}
\section{Fermi points in the Chiral-LL phase}\label{SMSec:I}
In the quantum Ising model with Dzyaloshinskii-Moriya (DM) interaction, the dispersion relation is known as
$\epsilon_k = 2\big(\sqrt{1+g^2-2g\cos k} + D\sin k\big)$ (for example, see Ref.~\onlinecite{SMSoltania2019}).
The energy gap $\Delta = \max\big(\min(\epsilon_k), 0\big)$ is determined by the condition $d\epsilon_k/dk = 0$ ($k < 0$).
In the FM phase where $0 < D, g < 1$, the energy gap $\Delta$ goes down gradually with the increase of $D$,
and vanishes as $2\sqrt{1-g^2}(D_c-D)$ in the vicinity of $D_c = 1$.
In the transition line of $D_c = 1$, the momentum locates at $k_F = -\cos^{-1}(g)$.

\begin{figure}[!ht]
\centering
\includegraphics[width=0.45\columnwidth, clip]{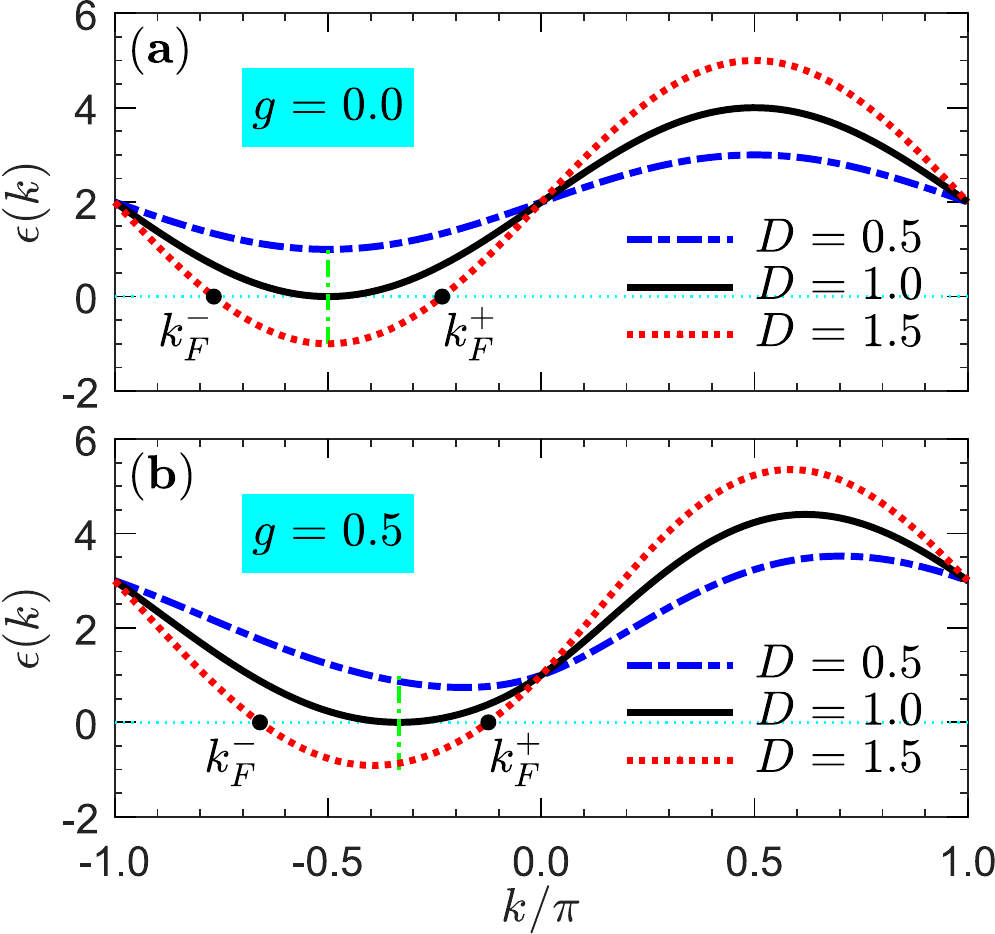}\\
\caption{(a) Energy spectrum plotted as a function of momentum $k$ for the DM interaction $D$ = 0.5 (blue dot-dashed line), $D$ = 1.0 (black solid line), and $D$ = 1.5 (red dotted line)
  at fixed magnetic field $g$ = 0.0. The Fermi points $k_F^{-}$ and $k_F^{+}$ satisfying $\epsilon_k = 0$ are marked by two black points.
  (b) The same as (a) but for $g$ = 0.5.
  }\label{FIGSM-QIMFS}
\end{figure}

As the ground state becomes the chiral-LL phase, there are two Fermi points which are given by \cite{SMSoltania2019}
\begin{equation}
k_F^{-} = -\cos^{-1}\left[ \frac{g}{D^2} - \sqrt{\left(1-\frac{1}{D^2}\right)\left(1-\frac{g^2}{D^2}\right)} \right]
\end{equation}
and
\begin{equation}
k_F^{+} = -\cos^{-1}\left[ \frac{g}{D^2} + \sqrt{\left(1-\frac{1}{D^2}\right)\left(1-\frac{g^2}{D^2}\right)} \right].
\end{equation}
It could be verified that $k_F^{-} + k_F^{-} = -\pi$.
Figure~\ref{FIGSM-QIMFS} shows the energy spectrum plotted as a function of momentum $k$ for different DM interaction $D$ with fixed (a) $g = 0.0$ and (b) $g = 0.5$, respectively.

\section{Low- and high-temperature behaviors of GR at the Ising transition}\label{SMSec:II}

At the Ising transition where $g = 1$, the dispersion relation reads $\epsilon_k = 4\big(\left\vert\sin\frac{k}{2}\right\vert + \sin\frac{k}{2}\cos\frac{k}{2}\big)$.
We have $\epsilon_k = 4\sin\frac{k}{2} \big(1+D\cos\frac{k}{2}\big)$ when $k\in[0, \pi]$ and
$\epsilon_k = -4\sin\frac{k}{2} \big(1-D\cos\frac{k}{2}\big)$ when $k\in[-\pi,0]$.
It could checked that the dispersion relation remains unchanged under the simultaneous inversion operations $k \to -k$ and $D \to -D$.
The magnetic expansion coefficient of Eq.~(\textcolor{red}{9}) in the main text is reduced to $\alpha_g = \alpha_g(k>0) + \alpha_g(k<0)$ with
\begin{align}\label{SMEQ:MagExpHc}
\alpha_g(k>0)
&= \frac{1}{2\pi}\int_{0}^{\pi} dk \beta^2(1-\cos k)\Big(1+D\cos\frac{k}{2}\Big) \textrm{sech}^2\left( 2\beta\sin\frac{k}{2}\Big(1+D\cos\frac{k}{2}\Big) \right) \nonumber\\
&= \frac{1}{2\pi}\int_{0}^{\pi/2} dk (2\beta)^2 \sin^2k (1+D\cos k) \textrm{sech}^2\big( 2\beta\sin k(1+D\cos k) \big).
\end{align}
Similarly, the specific heat of Eq.~(\textcolor{red}{6}) in the main text is $C_v = C_v(k>0) + C_v(k<0)$ with
\begin{align}\label{SMEQ:CvHc}
C_v(k>0) &= \frac{1}{2\pi}\int_{0}^{\pi} dk \Big(\frac{\beta\epsilon_k}{2}\Big)^2 \textrm{sech}^2\Big(\frac{\beta\epsilon_k}{2}\Big)  \nonumber\\
&= \frac{1}{2\pi}\int_{0}^{\pi} dk \left( 2\beta\sin\frac{k}{2}\Big(1+D\cos\frac{k}{2}\Big) \right)^2 \textrm{sech}^2\left( 2\beta\sin\frac{k}{2}\Big(1+D\cos\frac{k}{2}\Big) \right)  \nonumber\\
&= \frac{2}{2\pi}\int_{0}^{\pi/2} dk \Big(2\beta\sin k (1+D\cos k)\Big)^2 \textrm{sech}^2\Big( 2\beta\sin k (1+D\cos k) \Big).
\end{align}
Let us introduce an auxiliary integral (which is Eq.~(\textcolor{red}{11}) in the main text)
\begin{align}\label{SMEQ:FuncInu}
\mathcal{I}_{\upsilon}(T, D)
=& \frac{\upsilon}{\pi} \int_0^{\pi/2} dk (1+D\cos k)^{\upsilon} (2\beta\sin k)^2 \textrm{sech}^2 \big[2\beta\sin k (1+D\cos k)\big]
\end{align}
where $\upsilon$ represents the order of this integral.
It is easily found that Eq.~\eqref{SMEQ:MagExpHc} and Eq.~\eqref{SMEQ:CvHc} could be expressed by $\mathcal{I}_{\upsilon}(T, D)$, and the GR is
\begin{equation}\label{SMEQ:GruneisenRatio}
\Gamma(T, g_c) = -\frac{\alpha_g}{C_v} = \frac{[\mathcal{I}_1(T, D) + \mathcal{I}_1(T, -D)]/2}{[\mathcal{I}_2(T, D) + \mathcal{I}_2(T, -D)]/2}
= \frac{\mathcal{I}_1(T, D) + \mathcal{I}_1(T, -D)}{\mathcal{I}_2(T, D) + \mathcal{I}_2(T, -D)}.
\end{equation}

\subsection{high-temperature behavior}
In the high-temperature situation where $T \to \infty$ (i.e., $\beta \to 0$), $\textrm{sech}^2(x) \to 1$ (as $x\to0$) and the auxiliary integral in Eq.~\eqref{SMEQ:FuncInu} is approximately
\begin{align}
\mathcal{I}_{\upsilon}(T, D) \simeq& \frac{\upsilon}{\pi} \int_0^{\pi/2} dk (1+D\cos k)^{\upsilon} (2\beta\sin k)^2
= \frac{4\upsilon\beta^2}{\pi} \int_0^{\pi/2} dk (1+D\cos k)^{\upsilon} \sin^2k.
\end{align}
If $\upsilon = 1$,
\begin{align}
\mathcal{I}_{1}(T, D) = \frac{4\beta^2}{\pi} \int_0^{\pi/2} dk (1+D\cos k) \sin^2k = \frac{4\beta^2}{\pi}\left(\frac{D}{3}+\frac{\pi}{4}\right) = \beta^2\left(\frac{4D}{3\pi}+1\right).
\end{align}
If $\upsilon = 2$,
\begin{align}
\mathcal{I}_{2}(T, D) = \frac{8\beta^2}{\pi} \int_0^{\pi/2} dk (1+D\cos k)^{2} \sin^2k = \frac{8\beta^2}{\pi}\left(\frac{\pi(4+D^2)}{16}+\frac{2D}{3}\right) = \beta^2\left(\frac{4+D^2}{2}+\frac{16D}{3\pi}\right).
\end{align}
The GR is calculated as
\begin{equation}\label{SMEQ:GruneisenRatio}
\Gamma(T, g_c) = \frac{\mathcal{I}_1(T, D) + \mathcal{I}_1(T, -D)}{\mathcal{I}_2(T, D) + \mathcal{I}_2(T, -D)}
= \frac{2}{\frac{4+D^2}{2}\times2} = \frac{2}{4+D^2} \leq \frac12.
\end{equation}

\subsection{low-temperature behavior}
In the low-temperature situation where $T \to 0$ (i.e., $\beta \to \infty$), $\textrm{sech}^2(x) \simeq 4e^{-2x}$ (as $x\gg1$) and the auxiliary integral in Eq.~\eqref{SMEQ:FuncInu} is approximately
\begin{align}
\mathcal{I}_{\upsilon}(T, D) \simeq& \frac{\upsilon}{\pi} \int_0^{\pi/2} dk (1+D\cos k)^{\upsilon} (4\beta\sin k)^2 e^{-4\beta \sin k(1+D\cos k)}.
\end{align}
If $\upsilon = 1$,
\begin{align}
\mathcal{I}_{1}(T, D) &= \frac{1}{\pi} \int_0^{\pi/2} dk (1+D\cos k) (4\beta\sin k)^2 e^{-4\beta \sin k(1+D\cos k)} \nonumber\\
&= -\frac{4\beta^2}{\pi} \frac{\partial}{\partial \beta} \Big[ \int_0^{\pi/2} dk \sin k e^{-4\beta \sin k(1+D\cos k)} \Big] \nonumber\\
&= -\frac{4\beta^2}{\pi} \frac{\partial}{\partial \beta} \Big[ \int_0^1 dt e^{-4\beta\sqrt{1-t^2}(1+Dt)} \Big] \nonumber\\
&= -\frac{4\beta^2}{\pi} \frac{\partial}{\partial \beta} \Big[ \frac{1}{(1+D)^2(4\beta)^2} \Big] \nonumber\\
&= \frac{1}{2\pi\beta(1+D)^2}.
\end{align}
If $\upsilon = 2$,
\begin{align}
\mathcal{I}_{2}(T, D) &= \frac{2}{\pi} \int_0^{\pi/2} dk (1+D\cos k)^2 (4\beta\sin k)^2 e^{-4\beta \sin k(1+D\cos k)} \nonumber\\
&= \frac{2(4\beta)^2}{\pi} \frac{\partial^2}{\partial (4\beta)^2} \Big[ \int_0^{\pi/2} dk e^{-4\beta \sin k(1+D\cos k)} \Big] \nonumber\\
&= \frac{2(4\beta)^2}{\pi} \frac{\partial^2}{\partial (4\beta)^2} \Big[ \frac{1}{4\beta(1+D)} \Big] \nonumber\\
&= \frac{1}{\pi\beta(1+D)}.
\end{align}
The GR is calculated as
\begin{equation}\label{SMEQ:GruneisenRatio}
\Gamma(T, g_c) = \frac{\mathcal{I}_1(T, D) + \mathcal{I}_1(T, -D)}{\mathcal{I}_2(T, D) + \mathcal{I}_2(T, -D)}
= \frac{\frac{1}{(1+D)^2}+\frac{1}{(1-D)^2}}{2\big[\frac{1}{1+D} + \frac{1}{1-D}\big]} = \frac{1+D^2}{2(1-D^2)} \geq \frac12.
\end{equation}

Finally, we present the low-$T$ behavior of the specific heat. According to the formalization above, it is known that
\begin{equation}
C_v(T, D) = \frac{1}{2}[\mathcal{I}_2(T, D) + \mathcal{I}_2(T, -D)] = \frac{1}{2} \Big(\frac{1}{1+D} + \frac{1}{1-D}\Big)\frac{1}{\pi\beta} = \frac{k_BT}{\pi(1-D^2)}.
\end{equation}
However, for the transverse-field Ising model ($D = 0$), the specific heat is known to be \cite{SMBloteCN1986,SMAffleck1986}
\begin{equation}
C_v(T, D = 0) =\frac{\pi}{12} k_BT \neq \frac{1}{\pi} k_BT.
\end{equation}
The discrepancy in the coefficient comes from the approximation $\textrm{sech}^2(x) \sim 4e^{-2x}$.
After correcting the prefactor, we conclude that the specific heat in the low-temperature regime behaves as
\begin{equation}\label{SMEQ:CvLowT}
C_v(T, D) = \frac{\pi}{12(1-D^2)} k_BT.
\end{equation}
This formula Eq.~\eqref{SMEQ:CvLowT} has also been checked numerically and it is inferred that the scaled free energy coefficient
$\Psi_g(T, D) = \frac{\pi}{12(1-D^2)}$ (cf. Fig.~\textcolor{red}{2} in the main text).

\section{The nondivergence of the GR in the quantum critical region}\label{SMSec:III}

In the neighbor of the generic quantum critical point, the singular part of the free energy has the following scaling behavior \cite{SMZGRS2003}
\begin{equation}\label{SMEQ:FreeEgPsi}
F_s(T) \propto (T/T_0)^{(d+z)/z} \Psi\left(\frac{r}{(T/T_0)^{1/z\nu}}\right),
\end{equation}
where $r = (g-g_c)/g_c$ and $T_0$ is a nonuniversal characteristic temperature scale.
$z$ and $\nu$ ($z = \nu = 1$) are the critical exponents, and $\Psi(x)$ is the smooth hyperscaling function.
In the quantum critical region where $|r|/(T/T_0)^{1/z\nu} \ll 1$,
the GR near the quantum critical point has the following form \cite{SMZGRS2003}
\begin{equation}
\Gamma(T, r\to0) = - G_T T^{-1/(z\nu)}
\end{equation}
where
\begin{equation}\label{SMEQ:GTPsi}
G_T = \frac{(d+z-1/\nu)z\Psi'(0)}{d(d+z)\Psi(0)}\frac{T_0^{1/z\nu}}{g_c}.
\end{equation}
Generally, the linear tem in the scaling expansion series of Eq.~\eqref{SMEQ:FreeEgPsi} in powers of $r/(T/T_0)^{1/z\nu}$ is nonzero (i.e., $\Psi'(0) \neq 0$ as inferred from Eq.~\eqref{SMEQ:GTPsi}), and thus the GR diverges at the critical point.
However, in the Ising transition of the model under consideration, the linear scaling term vanishes and the quadratic one serves as the major contribution,
giving rise to a scarce constant GR \cite{SMWuZhuSi2018,SMZhang2019}.

In what follows we shall demonstrate that the linear scaling term in $\Psi(x)$ vanishes in the transverse-field Ising model with DM interaction.
Following the recipe depicted in Ref.~\onlinecite{SMWuZhuSi2018}, we introduce a crossover momentum $k_c \simeq T$ such that
when $k < k_c$, $\sin(k/2)/T \ll 1$ and when $k > k_c$, $\sin(k/2)/T \gg 1$.
Consequently, we split the integration interval of the free energy $F(T)$ into three regions,
\begin{equation}\label{SMEQ:FreeEg3Part}
F(T) = -\frac{1}{2\pi\beta} \left(\int_{-\pi}^{-k_c} + \int_{-k_c}^{k_c} + \int_{k_c}^{\pi}\right) dk \ln(e^{\mathcal{A}} + e^{-\mathcal{A}}),
\end{equation}
with $\mathcal{A} \equiv \beta\epsilon_k/2 = \beta\Big(\sqrt{(1-g)^2+4g\sin^2(k/2)} + D\sin k\Big)$.
When $\mathcal{A} \ll 1$, $\ln(e^{\mathcal{A}} + e^{-\mathcal{A}}) \simeq \ln2 + \mathcal{A}^2/2 + \mathcal{O}(\mathcal{A}^2)$. Hence,
\begin{align}\label{SMEQ:FreeEgLowT}
F(T) &\simeq -\frac{1}{2\pi\beta} \int_{-k_c}^{k_c} dk \Big(\ln2 + \frac{\mathcal{A}^2}{2}\Big) \nonumber\\
&= -\frac{1}{2\pi\beta} \int_{-k_c}^{k_c} dk \Big\{\ln2 + \frac{\beta^2}{2}\big[(1-g)^2+4g\sin^2(k/2) + D^2\sin^2k\big]\Big\},
\end{align}
where in the last step we have used the fact that the integration of an \textit{odd} function is zero.
With $k_c \simeq T$ and $\cos k_c \simeq 1$ in mind, the free energy near the critical point in the low-temperature limit reads
\begin{align}\label{SMEQ:FreeEgLowTFnl}
F(T) = -\frac{T^2}{\pi} \Big[a_{g,D} + \frac12 \Big(\frac{1-g}{T}\Big)^2\Big]
\end{align}
where $a_{g,D} = \ln2 + (g+D^2/4)/6$.
It is found from Eq.~\eqref{SMEQ:FreeEgLowTFnl} that the linear scaling term of $(1-g)/T$ indeed vanishes and thus $\Psi'(0) = 0$ for arbitrary $D$.
This in turn accounts for the nondivergence of the GR along the Ising transition line.
More precisely, by noticing that
entropy $S = -(\partial F/\partial T)$,
magnetic expansion coefficient $\alpha = -(\partial S/\partial g)_T$,
and specific heat $C_v = T(\partial S/\partial T)_g$,
we obtain that
\begin{equation}
S \simeq \Big(2\ln 2 + \frac{g+D^2/4}{3}\Big)T \Rightarrow \left\{
\begin{array}{l}
 \alpha  \simeq T/3 \\
 {C_v} \simeq \big((g+D^2/4)/3 + 2\ln 2\big)T \\
 \end{array}  \right.,
\end{equation}
and thus the GR at the critical point is
\begin{equation}\label{ss1}
 \Gamma  = \frac{\alpha }{{{C_v}}} \sim {\rm const}.
\end{equation}

\section{GR in the zero-temperature limit: The Ising transition as an example}\label{SMSec:IV}

For the transverse-field Ising model (i.e., $D = 0$), the GR in the low-temperature limit is known to obey the following behavior \cite{SMWuZhuSi2018}
\begin{align}\label{SMEQ:GRTFIMZeroTDEQ0}
\left\{
\begin{array}{l}
\vspace{0.20cm}
\Gamma(T\to0, g =    g_c) = \frac12 \\
\Gamma(T\to0, g \neq g_c) = \frac{C}{g-g_c}
\end{array} \right.
\end{align}
Here, $C (= 1)$ is a constant. The quantum critical point is identified as a discontinuity point of the second kind.
On the other hand, for arbitrary $D$, the GR $\Gamma(T, g)$ is found to be
\begin{align}\label{SMEQ:GRTFIMZeroTDNeq0}
\left\{
\begin{array}{l}
\vspace{0.20cm}
\Gamma(T\to0, g =    g_c) = \frac{1+D^2}{2(1-D^2)} \\
\Gamma(T\to0, g \neq g_c) = \frac{C_{\pm}}{g-g_c}
\end{array} \right.
\end{align}
where $C_{+}$ ($C_{-}$) stands for the coefficient when $r>0$ ($r<0$) with $r=(g-g_c)/g_c$, and $C_{+} \neq C_{-}$.
Clearly, Eq.~\eqref{SMEQ:GRTFIMZeroTDEQ0} is nothing but a special case of Eq.~\eqref{SMEQ:GRTFIMZeroTDNeq0}.
However, the closed form of $C_{\pm}$ is still lacking. Numerically, $C_{+}$ is slightly larger than 1 while $C_{-}$ is slightly smaller than 1.

\begin{figure}[!ht]
\centering
\includegraphics[width=0.80\columnwidth, clip]{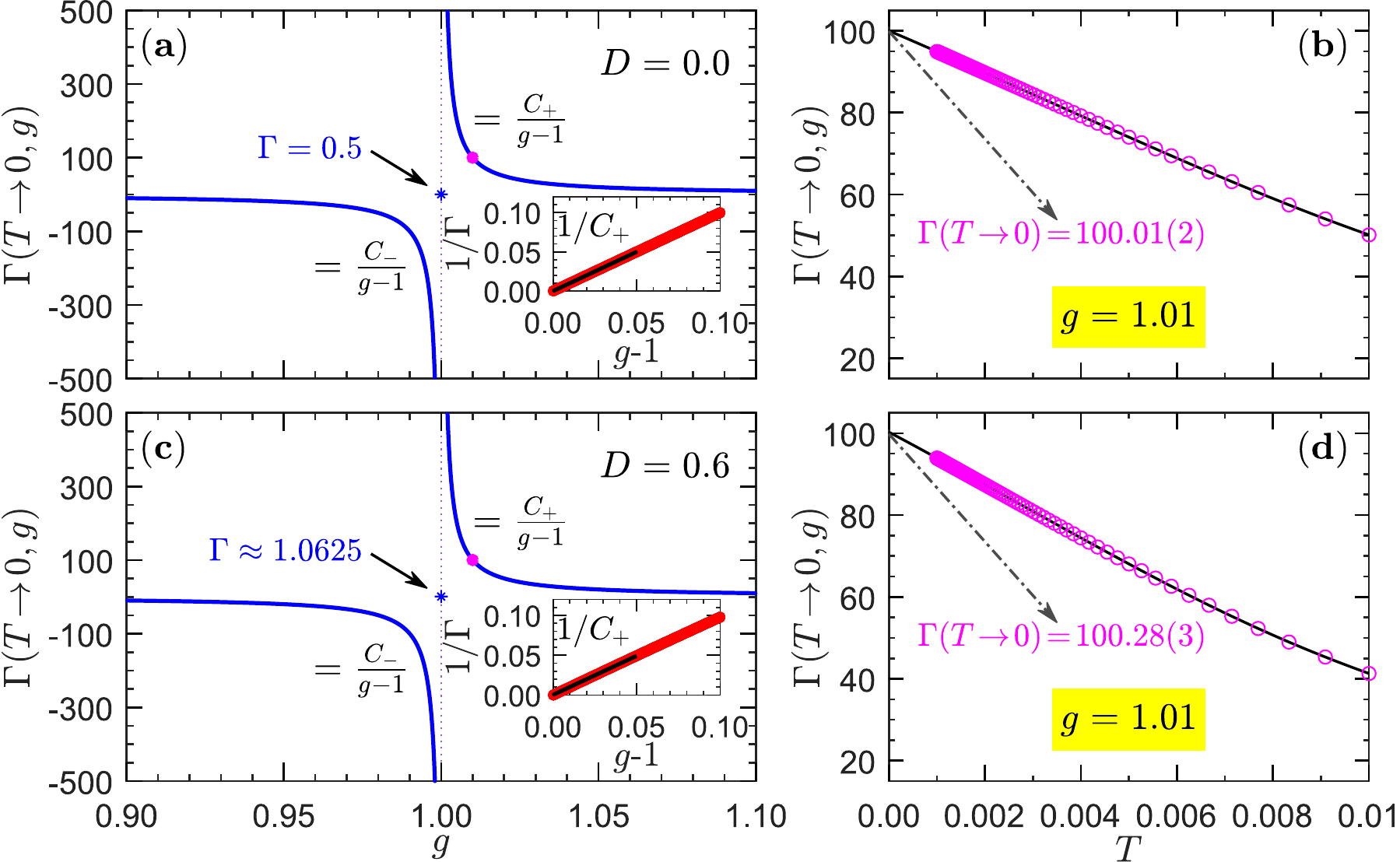}\\
\caption{(a) The magnetic field $g$ dependence of the GR $\Gamma(T\to0, g)$ in the zero temperature with DM interaction $D$ = 0.0.
  Inset: The fitting of the coefficient $C_+$, yielding a value of 1.00. The red symbols are the data points while the black line represents the linear fitting.
  (b) Extrapolation of the GR $\Gamma(T, g)$ at fixed $g = 1.01$ w.r.t. the temperature.
  The extrapolated value is 100.01(2), which is consistent with the value shown in panel (a) (marked by a pink circle).
  The panels (c) and (d) are respectively analogs of (a) and (b) but for $D$ = 0.6.
  In the inset of panel (c), $C_{+} \approx 1.02$.
  }\label{FIGSM-GRZeroT}
\end{figure}

In the zero-temperature limit, the GR near the transition point is determined by the energy gap $\Delta(g)$ \cite{SMBachusKT2020},
\begin{align}\label{SMEQ:GRIsingDef}
\lim_{T\to0^{+}}\Gamma(T, g) = \frac{\Delta'(g)}{\Delta(g)},
\end{align}
where the numerator represents the derivative of $\Delta(g)$.
The energy gap $\Delta(g) = \max\big(\min(\epsilon_k), 0\big)$ is determined by the condition
$d\epsilon_k/dk = 2g\sin k/\sqrt{1+g^2-2g\cos k} + 2D\cos k = 0$ ($k_o \leq 0$),
where the dispersion relation is $\epsilon_k = 2\big(\sqrt{1+g^2-2g\cos k} + D\sin k\big)$.
When $D = 0$, $k_o = 0$ and $\Delta(g) = 2|1-g|$.
According to Eq.~\eqref{SMEQ:GRIsingDef} we have $\lim_{T\to0^{+}}\Gamma(T, g) = 1/(g-1)$, implying that $C = C_{\pm} = 1$.
This is in accordance with the results shown in Fig.~\ref{FIGSM-GRZeroT}(a) and Fig.~\ref{FIGSM-GRZeroT}(b).
However, when $D > 0$ the value of $k_o$ is intricate and the explicit expression of $C_{\pm}$ is unclear.
Taking $D = 0.6$ as an example, Fig.~\ref{FIGSM-GRZeroT}(c) indicates that $C_+ \approx 1.02$, which are close to 1.
We note in passing that $C_- \approx 0.96$, which is slightly smaller than 1.

\section{Divergent GR at the first-order transition: $g$ = 0 as an example}\label{SMSec:V}

In this section we present the low-temperature behavior of the GR at the first-order FM--chiral-LL transition.
For simplicity, we focus on the zero magnetic field case with $g$ = 0.
The dispersion relation is thus simplified as $\epsilon_k = 2(1+\sin k\big)$.
The magnetic expansion coefficient of Eq.~(\textcolor{red}{17}) in the main text is reduced to
\begin{align}
\alpha_g
&= \frac{1}{2\pi}\int_{-\pi}^{\pi} dk (\beta\sin k) \big(\beta(1+\sin k)\big) \textrm{sech}^2 \big(\beta(1+\sin k)\big) \nonumber\\
&= -\frac{\beta^2}{\pi}\int_{0}^{\pi/2} dk \sin k(1-\sin k) \textrm{sech}^2 \big(\beta(1-\sin k)\big).
\end{align}
Let $t = 1-\sin k$, then $k = \arcsin(1-t)$ and $dk = -\frac{dt}{\sqrt{t(2-t)}}$. Therefore,
\begin{align}
\alpha_g
&= -\frac{\beta^2}{\pi} \int_{0}^{1} \frac{dt}{\sqrt{t(2-t)}} t(1-t) \textrm{sech}^2 \big(\beta t\big) \nonumber\\
&\simeq -\frac{\beta^2}{\sqrt2\pi} \int_{0}^{1} dt \sqrt{t}(1-t) \textrm{sech}^2 \big(\beta t\big) \nonumber\\
&\simeq -\frac{\beta^2}{\sqrt2\pi} \int_{0}^{1} dt \sqrt{t} \textrm{sech}^2 \big(\beta t\big) \nonumber\\
&\simeq -\frac{\sqrt{\beta}}{\sqrt2\pi} \left[\int_{0}^{1} d(\beta t) \sqrt{\beta t} \textrm{sech}^2 \big(\beta t\big)\right] \nonumber\\
&= -\frac{\sqrt{\beta}}{\sqrt2\pi} \mathcal{F}(1/2).
\end{align}
Here, the parameter-dependent integral
\begin{align}
\mathcal{F}(\mu) = \int_0^{\infty} dx x^{\mu} \textrm{sech}^2(x) = 2^{1-\mu} (1-2^{1-\mu}) \Gamma(1+\mu)\zeta(\mu),
\end{align}
where $\zeta(\mu)$ is the Riemann $\zeta$-function and $\Gamma(1+\mu) = \mu\Gamma(\mu)$. Specifically, we have
$\mathcal{F}(1/2) = \frac{(1-\sqrt2)\sqrt{2\pi}}{2}\zeta(1/2)$ and
$\mathcal{F}(3/2) = \frac{3(\sqrt2-1)\sqrt{\pi}}{8}\zeta(3/2)$.
Hence, in the low-temperature regime the magnetic expansion coefficient
\begin{equation}
\alpha_D = \frac{(\sqrt2-1)\zeta(\frac12)}{2\sqrt{\pi T}}.
\end{equation}

On the other hand, the specific heat of Eq.~(\textcolor{red}{6}) in the main text is
\begin{align}
C_v
&= \frac{1}{2\pi}\int_{-\pi}^{\pi} dk \big(\beta(1+\sin k)\big)^2 \textrm{sech}^2 \big(\beta(1+\sin k)\big) \nonumber\\
&= \frac{1}{\pi}\int_{0}^{\pi/2} dk \big(\beta(1-\sin k)\big)^2 \textrm{sech}^2 \big(\beta(1-\sin k)\big).
\end{align}
Let $t = 1-\sin k$, then $k = \arcsin(1-t)$ and $dk = -\frac{dt}{\sqrt{t(2-t)}}$. Therefore,
\begin{align}
C_v
&= \frac{\beta^2}{\pi}\int_{0}^{1} dt \frac{t^{3/2}}{\sqrt{2-t}} \textrm{sech}^2 \big(\beta t\big) \nonumber\\
&\simeq \frac{\beta^2}{\sqrt2 \pi}\int_{0}^{1} dt t^{3/2} \textrm{sech}^2 \big(\beta t\big) \nonumber\\
&= \frac{1}{\pi\sqrt{2\beta}} \int_{0}^{1} d(\beta t) (\beta t)^{3/2} \textrm{sech}^2 \big(\beta t\big) \nonumber\\
&= \frac{1}{\pi\sqrt{2\beta}} \mathcal{F}(3/2)  \nonumber\\
&= \frac{3(\sqrt2-1)\zeta(\frac32)}{8\sqrt{2\pi}} \sqrt{T}
\end{align}

The GR $\Gamma(T\to0, D_c) = -\alpha_D/C_v$ at the critical point is presented in Eq.~(\textcolor{red}{20}) in the main text.

\section{GR in the zero-temperature limit: $g$ = 0 as an example}\label{SMSec:VI}

For the first-order FM--chiral-LL transition ($g = 0$), the GR in the zero-temperature limit is found to be $\Gamma(T\to0, D) = 1/(1-D)$
(cf. Eq.~(\textcolor{red}{18}) in the main text). Here we present a direct but more involved derivation of this relation.
This derivation is composed of two steps, i.e., the derivation of the low-temperature behaviors of specific heat and magnetic expansion coefficient.

\subsection{The specific heat in the low-temperature limit}
We begin by calculating the low-temperature behavior of the specific heat. With $g$ = 0 in mind we have
\begin{align}
C_v &= \frac{1}{2\pi}\int_{-\pi}^{\pi} dk \big[\beta(1+D\sin k)\big]^2 \textrm{sech}^2 \big[\beta(1+D\sin k)\big] \nonumber\\
&\simeq \frac{1}{2\pi}\int_{-\pi}^{0} dk \big[\beta(1+D\sin k)\big]^2 \textrm{sech}^2 \big[\beta(1+D\sin k)\big] \nonumber\\
&\simeq \frac{4\beta^2}{2\pi}\int_{-\pi}^{0} dk (1+D\sin k)^2 e^{-2\beta(1+D\sin k)} \nonumber\\
&= \frac{4\beta^2}{\pi}\int_{0}^{\pi/2} dk (1-D\sin k)^2 e^{-2\beta(1-D\sin k)} \nonumber\\
&= \frac{4\beta^2e^{-2\beta}}{\pi}\Big[ \int_{0}^{\pi/2} dk e^{2\beta D\sin k} - 2D\int_{0}^{\pi/2} dk \sin k e^{2\beta D\sin k} + D^2 \int_{0}^{\pi/2} dk \sin^2 k e^{2\beta D\sin k} \Big].
\end{align}
Noticing that
\begin{align}
& \int_{0}^{\pi/2} dk \sin k e^{2\beta D\sin k} = \frac{1}{2D}\frac{\partial}{\partial\beta} \int_{0}^{\pi/2} dk e^{2\beta D\sin k}, \nonumber\\
& \int_{0}^{\pi/2} dk \sin^2 k e^{2\beta D\sin k} = \frac{1}{(2D)^2}\frac{\partial^2}{\partial\beta^2} \int_{0}^{\pi/2} dk e^{2\beta D\sin k},
\end{align}
we have
\begin{align}
C_v &= \frac{4\beta^2e^{-2\beta}}{\pi}\Big[ \mathcal{R}_D(\beta) - \frac{\partial}{\partial\beta}\mathcal{R}_D(\beta) + \frac14 \frac{\partial^2}{\partial\beta^2} \mathcal{R}_D(\beta) \Big]
\end{align}
where
\begin{align}
\mathcal{R}_D(\beta) = \int_{0}^{\pi/2} dk e^{2\beta D\sin k} = \frac{\pi}{2}\big[I_0(2\beta D) + L_0(2\beta D) \big].
\end{align}
Here, $I_0(z)$ is the zero-order modified Bessel function of the first kind while $L_0(z)$ is the zero-order modified Struve function.
According to the recurrence relations of $I_n(z)$, we have \cite{SMAbramowitzStegun1972}
\begin{align}
I'_n(z) = I_{n-1}(z) - \frac{n}{z} I_n(x)
\end{align}
and
\begin{align}
I''_n(z) = I_{n-2}(z) - \frac{2n-1}{z} I_{n-1}(x) + \frac{n(n+1)}{z^2} I_{n}(x),
\end{align}
and similar relations hold for the $L_n(z)$.
Further, for large $z$, i.e., fixed $n$ and $z \gg n$, $I_n(z) \simeq \frac{e^{z}}{\sqrt{2\pi z}}$ and
$L_{n}(z) \simeq I_{-n}(z) - \frac{(z/2)^{n-1}}{\sqrt{\pi}\Gamma(n+1/2)}$.
This means that in the large $z$ limit we have
$L_{n}(z) \simeq I_n(z) \simeq \frac{e^{z}}{\sqrt{2\pi z}}$,
which is irrelevant to $n$.
Taken together, in the low-temperature limit where $\beta\to\infty$, we have
\begin{align}
C_v &= 4\beta^2e^{-2\beta} \Big[ I_0(2\beta D) - \frac{\partial}{\partial\beta} I_0(2\beta D) + \frac14 \frac{\partial^2}{\partial\beta^2} I_0(2\beta D) \Big].
\end{align}
Below we will simplify the expression in the bracket. By using of the recurrence relations we have
\begin{eqnarray}
\tilde{\mathcal{F}}_D(\beta)
&=& I_0(2\beta D) - \frac{\partial}{\partial\beta} I_0(2\beta D) + \frac14 \frac{\partial^2}{\partial\beta^2} I_0(2\beta D) \nonumber\\
&=& I_0(2\beta D) - \left[I_{-1}(2\beta D) - \frac{0}{2\beta D} I_0(2\beta D)\right](2D) \nonumber\\
&& + \frac{1}{4} \left[I_{-2}(2\beta D) - \frac{-1}{2\beta D} I_{-1}(2\beta D) + \frac{0}{(2\beta D)^2}I(2\beta D) \right](2D)^2 \nonumber\\
&=& I_0(2\beta D) + \left[\frac{D}{2\beta^2} - 2D\right] I_{-1}(2\beta D) + D^2 I_{-2}(2\beta D) \nonumber\\
&\simeq& (1-D)^2 I_0(2\beta D) \simeq \frac{(1-D)^2}{\sqrt{4\pi\beta D}} e^{2\beta D}.
\end{eqnarray}
Thus, the asymptotical behavior of the specific heat is
\begin{align}
C_v \simeq \frac{2\beta^2(1-D)^2}{\sqrt{\pi\beta D}} e^{-2\beta(1-D)}.
\end{align}

\subsection{The magnetic expansion coefficient in the low-temperature limit}
We proceed to calculate the low-temperature behavior of the magnetic expansion coefficient. Likely, we have
\begin{align}
\alpha_T &= \frac{1}{2\pi}\int_{-\pi}^{\pi} dk \beta^2 \sin k(1+D\sin k) \textrm{sech}^2 \big[\beta(1+D\sin k)\big] \nonumber\\
&\simeq \frac{1}{2\pi}\int_{-\pi}^{0} dk \beta^2 \sin k(1+D\sin k) \textrm{sech}^2 \big[\beta(1+D\sin k)\big] \nonumber\\
&\simeq \frac{4\beta^2}{2\pi}\int_{-\pi}^{0} dk \sin k(1+D\sin k) e^{-2\beta(1+D\sin k)} \nonumber\\
&= -\frac{4\beta^2e^{-2\beta}}{\pi}\int_{0}^{\pi/2} dk \sin k(1-D\sin k) e^{2\beta D\sin k} \nonumber\\
&= -\frac{4\beta^2e^{-2\beta}}{\pi} \frac{1}{2D} \Big[ \frac{\partial}{\partial\beta}\mathcal{R}_D(\beta) - \frac12 \frac{\partial^2}{\partial\beta^2} \mathcal{R}_D(\beta) \Big] \nonumber\\
&= -\frac{2\beta^2e^{-2\beta}}{D} \Big[ \frac{\partial}{\partial\beta}I_0(2\beta D) - \frac12 \frac{\partial^2}{\partial\beta^2} I_0(2\beta D) \Big].
\end{align}
The expression in the bracket could be simplified as
\begin{eqnarray}
\tilde{\mathcal{F}}_D(\beta)
&=& \frac{\partial}{\partial\beta} I_0(2\beta D) - \frac12 \frac{\partial^2}{\partial\beta^2} I_0(2\beta D) \nonumber\\
&=& \left[I_{-1}(2\beta D) - \frac{0}{2\beta D} I_0(2\beta D)\right](2D) - \frac{1}{2} \left[I_{-2}(2\beta D) - \frac{-1}{2\beta D} I_{-1}(2\beta D) + \frac{0}{(2\beta D)^2}I(2\beta D) \right](2D)^2 \nonumber\\
&=& \left[2D + \frac{D}{\beta}\right] I_{-1}(2\beta D) - 2D^2 I_{-2}(2\beta D) \nonumber\\
&\simeq& 2D(1-D) I_0(2\beta D) \simeq \frac{1-D}{\sqrt{\pi\beta D}} e^{2\beta D}.
\end{eqnarray}
Hence, asymptotical behavior of the magnetic expansion coefficient is
\begin{align}
\alpha_T \simeq -\frac{2\beta^2(1-D)}{\sqrt{\pi\beta D}} e^{-2\beta(1-D)}.
\end{align}

To summarize, we find that the GR in the zero-temperature limit behaviors as
\begin{align}
\Gamma(T\to0, D) = -\frac{\alpha_T}{C_v} = \frac{1}{1-D}.
\end{align}
As expected, this result is in accordance with the energy gap analysis shown in Eq.~(\textcolor{red}{18}) in the main text.

%


\end{document}